\documentstyle{elsart}
\begin{document}

\newcommand{\eqlabel}[1]{\label{eq.#1}}
\newcommand{\eq}[1]{(\ref{eq.#1})}	

\newcommand{\bra}[1]{\mbox{$\left\langle #1 \right|$}}
\newcommand{\ket}[1]{\mbox{$\left| #1 \right\rangle$}}
\newcommand{\braket}[2]{\mbox{$\left\langle #1 | #2 \right\rangle$}}
\newcommand{\av}[1]{\mbox{$\left| #1 \right|$}}
\newcommand{\ve}{\varepsilon}
\newcommand{\osc}{{\mbox{\rm \scriptsize osc}}}
\newcommand{\tot}{{\mbox{\rm \scriptsize tot}}}
\newcommand{\ground}{{\mbox{\rm \scriptsize ground}}}
\renewcommand{\max}{{\mbox{\rm \scriptsize max}}}
\renewcommand{\min}{{\mbox{\rm \scriptsize min}}}

\begin{frontmatter}
 \title{The maximum speed of dynamical evolution}

 \author{Norman Margolus\thanksref{nsf-darpa}}
 \address{Boston University Center for Computational
 Science\\ and MIT Artificial Intelligence Laboratory}

 \author{Lev B. Levitin}
 \address{Boston University Department of
                Electrical and Computer Engineering}
 \thanks[nsf-darpa]{Supported by NSF grant DMS-9596217 and
 by DARPA contract DABT63-95-C-0130}


 \begin{abstract}
 We discuss the problem of counting the maximum number of distinct
states that an isolated physical system can pass through in a given
period of time---its {\em maximum speed of dynamical evolution}.
Previous analyses have given bounds in terms of $\Delta E$, the
standard deviation of the energy of the system; here we give a strict
bound that depends only on $E-E_0$, the system's average energy minus
its ground state energy.  We also discuss bounds on information
processing rates implied by our bound on the speed of dynamical
evolution.  For example, adding one Joule of energy to a given
computer can never increase its processing rate by more than about
$3\times 10^{33}$ operations per second.
 \end{abstract}
 \end{frontmatter}

\section{Introduction}\label{intro}

In the realm of computation, the first two quantitative questions that
one is likely to ask about a machine are {\em (i) How much memory does
it have?} and {\em (ii) How fast does it run?}  In exploring the
computational limits of physical dynamics, one might try to ask the
same questions about an arbitrary physical system.

Question {\em (i)} essentially asks, ``How many distinct states can my
system be put into, subject to whatever physical constraints I have?''
This is really a very old question: the correct counting of physical
states is the problem that led to the introduction of Planck's
constant into physics\cite{planck}, and is the basis of all of quantum
statistical mechanics.  This question can be answered by a detailed
quantum mechanical counting of distinct (mutually orthogonal) states.
It can also be well approximated in the macroscopic
limit\cite{huang,wang} by simply computing the volume of phase space
accessible to the system, in units where Planck's constant is 1.

\smallskip

Question {\em (ii)} will be the focus of this paper.  This question
can be asked with various levels of sophistication.  Here we will
discuss a particularly simple measure of speed: the maximum number of
distinct states that the system can pass through, per unit of time.
For a classical computer, this would correspond to the maximum
number of operations per second.

For a quantum system, the notion of distinct states is well defined:
two states are distinct if they are orthogonal.  The connection
between orthogonality and rate of information processing has
previously been
discussed\cite{levitin,benioff,feynman,marg86,marg89,biafore}, but no
universal bound on computation rate was proposed.  The minimum time
needed for a quantum system to pass from one orthogonal state to
another has also previously been characterized, in terms of the
standard deviation of the energy $\Delta
E$\cite{mandelstam,peres,vaidman,pfeifer,braunstein}.  This bound
places no limit, however, on how fast a system with bounded average
energy can evolve (since $\Delta E$ can be arbitrarily large with
fixed $E$).  Bounds based directly on the average energy $E$ have
previously been proposed\cite{bremermann,bekenstein}, but these bounds
apply to the rate of communication of bits, rather than to the rate of
orthogonal evolution: difficulties associated with such bit-related
bounds are discussed in \cite{levitin}.  The new bounds derived in
this paper are also based on average energy, but they apply to rates
of orthogonal evolution.  For an ordinary macroscopic system, these
bounds are achievable: we show that adding energy increases the
maximum rate at which such a system can pass through a sequence of
mutually orthogonal states by a proportionate amount.

There has recently been much interest in the possibilities of {\em
quantum computers}: computers that can operate on superpositions of
computational states\cite{steane}.  Even isolated quantum computers
will, in general, pass through sequences of (nearly) mutually
orthogonal states in the course of their complicated time evolutions.
At the least, an efficient quantum computation should, for some
initial states, have a final state that is reasonably distinct from
its initial state.  Thus a bound on the rate of orthogonal evolution
is relevant in this case as well.

\clearpage

\section{Maximum rate of dynamics}\label{speed}

In the energy basis, quantum time evolutions are constructed out of
superpositions of frequency components.  One might expect from this
that, given a maximum energy eigenvalue, the frequency with
which states can change should be bounded by
 \begin{equation}\eqlabel{maxbound}
\nu_\perp \leq {E_\max \over h}
\end{equation}
 If we take our zero of energy at the ground state of the
system,\footnote{In all of this discussion, if for some reason no
energies below some $E_\min$ are allowed, we should take our zero of
energy there instead.} and consider long evolutions, then this
relation is true, as we will discuss below.  We will also show that,
given a fixed average energy $E$ (rather than a fixed maximum), there
is a similar bound
 \begin{equation}\eqlabel{longbound}
\nu_\perp \leq {2 E \over h},
\end{equation}
 where again we have taken our zero of energy at the ground state.
This equation has the following interpretation: in appropriate units,
the average energy of a macroscopic system is equal to the maximum
number of orthogonal states that the system can pass through per unit
of time.  This is the maximum rate that can be sustained for a long
evolution---the rate at which a system can oscillate between two
states is twice as great.

\subsection{Orthogonality time}\label{orthtime}

We begin our analysis by discussing the question of the minimum time
needed for {\em any} state of a given physical system to evolve into
an orthogonal state.

An arbitrary quantum state can be written as a superposition of energy
eigenstates
 \begin{equation}\eqlabel{super}
\ket{\psi_0} = \sum_n c_n \ket{E_n}
\end{equation}
 Here and throughout this paper we assume that our system has a
discrete spectrum, and that the states have been numbered so that the 
energy eigenvalues $\{E_n\}$ associated with the states
$\{\ket{E_n}\}$ are non-decreasing.  To simplify formulas, we will
choose our zero of energy so that $E_0=0$.

 Let $\tau_\perp$ be the time it takes for \ket{\psi_0} to evolve into
an orthogonal state.  We will now show that, with a fixed average
energy $E$, it is always true that 
 \begin{equation}\eqlabel{tauperp0}
\tau_\perp \geq {h \over 4 E}.
\end{equation}
  This result is somewhat surprising, since earlier results gave a
bound only in terms of $\Delta E$
 \begin{equation}\eqlabel{tauperpge}
\tau_\perp \geq {h \over 4 \Delta E}
\end{equation}
 This earlier bound would suggest that, given a fixed average energy,
one could construct a state with a very large $\Delta E$ in order to
achieve an arbitrarily short $\tau_\perp$.  We will show that this is
not the case.

\medskip
Let us begin by observing that
if \ket{\psi_0} is evolved for a time $t$ it becomes
 \begin{equation}\eqlabel{super1}
\ket{\psi_t} = \sum_n c_n e^{-i{E_n t \over \hbar}}\ket{E_n}
\end{equation}
Now we let
 \begin{equation}\eqlabel{orthog}
S(t) = \braket{\psi_0}{\psi_t} = \sum_{n=0}^\infty \av{c_n}^2 e^{-i{E_n t \over \hbar}} 
\end{equation}
 We want to find the smallest value of $t$ such that $S(t)$=0.  To do
this, we note that
 \newcommand{\eth}{{E_n t \over \hbar}}
\newcommand{\re}{\mbox{\rm Re}}
\newcommand{\im}{\mbox{\rm Im}}
 \begin{eqnarray}\eqlabel{reim}
\re(S) & = & \sum_{n=0}^\infty \av{c_n}^2 \cos(\eth) \nonumber\\
& \geq &
\sum_{n=0}^\infty \av{c_n}^2
\left( 1 - {2 \over \pi}(\eth + \sin(\eth))\right) \nonumber\\
& = & 1 - {2 E \over \pi \hbar}t + {2\over\pi}\im(S),
\end{eqnarray}
 where we have used the inequality $\cos x \geq 1 -
{2\over\pi}(x+\sin x)$, valid for $x\geq 0$.  But for any
value of $t$ for which $S(t)=0$, both $\re(S)=0$ and
$\im(S)=0$, and so Eq.~\eq{reim} becomes
 \begin{equation}
0 \geq 1- {4Et\over h}
\end{equation}
 Thus the earliest that $S(t)$ can possibly equal zero is when
$t=h/4E$, which proves Eq.~\eq{tauperp0}.  Of course Eq.~\eq{reim}
also gives approximately the same bound on how quickly we can have
approximate orthogonality, since if $\av{S(t)}$ is small, then so are
$\re(S)$ and $\im(S)$.


\smallskip

This bound is achievable if the spectrum of energies includes the
energy $2E$ (and is very nearly achievable if the spectrum includes a
value very close to this, as we would expect, for example, for any
ordinary macroscopic system).  In this case, we let
 \begin{equation}\eqlabel{fastest0}
\ket{\psi_0} = {\ket{0} + \ket{2E} \over \sqrt{2}}
\end{equation}
 which has average energy $E$.  This evolves in a time $t=h/4 E$ into
 \begin{equation}\eqlabel{fastest1}
\ket{\psi_t} = {\ket{0} - \ket{2E} \over \sqrt{2}}
\end{equation}
 which is orthogonal to \ket{\psi_0}.  If we evolve for the same
interval again, we will be back to \ket{\psi_0}: the evolution
oscillates between these two orthogonal states.  For these states,
$\Delta E = E$, and so both of the bounds Eq.~\eq{tauperp0} and
Eq.~\eq{tauperpge} are achieved.

There are also cases where Eq.~\eq{tauperp0} gives a much better bound
than Eq.~\eq{tauperpge}.  Consider, for example, the state
 \begin{equation}\eqlabel{bigdelta0}
\ket{\psi_0} \quad = \quad a\left(\ket{0} + \ket{\varepsilon}\right)
             \quad + \quad
               b\left(\ket{n\varepsilon} + \ket{(n+1)\varepsilon}\right)
\end{equation}
 which evolves into an orthogonal state in a time $\tau_\perp=h/2
\varepsilon$.  Given $E$, as long as we choose $\varepsilon < 2E$
(i.e., $\tau_\perp > h/4E$) the average energy of the first pair of
kets will be less than $E$.  Given $\varepsilon$, for large enough $n$
the average energy of the second pair of kets will be greater than
$E$.  Then we can always find coefficients $a$ and $b$ that make the
average energy of $\ket{\psi_0}$ be $E$ and also normalize the state.
But this state has a $\Delta E$ that depends on our choice of $n$: in
fact $\Delta E = \Theta(\sqrt{n})$.  With fixed $E$, $\Delta E$ can be
as large as we like.  Thus in this case, Eq.~\eq{tauperpge} is not a
useful bound and Eq.~\eq{tauperp0} is optimal.

\subsection{Cycles of orthogonal states}\label{cycles}

In the discussion above, we have seen that a quantum system with
average energy $E$ can be made to oscillate between two orthogonal
states with a frequency of $4E/h$.  Now we address the question of how
fast a quantum system can run through a long sequence of mutually
orthogonal states.  We begin by considering the very restricted set of
evolutions that pass through an exact cycle of $N$ mutually orthogonal
states at a constant rate.  In this case it is easy to show (see
Appendix~\ref{exact.orthog}) that
 \begin{equation}\eqlabel{tauofm}
 \tau_\perp\geq{N-1 \over N} {h \over 2E}
\end{equation}
 Thus for very long evolutions that form a closed cycle, the maximum
transition rate between orthogonal states is only half as great as it
is for an oscillation between two states.  In the next section, we
will show that this long-sequence asymptotic rate is achievable in
principle for any ordinary macroscopic system.  Here we will first
give an example of a system for which an {\em exact} cycle of $N$
mutually orthogonal states ({\em cf.} \cite{peg}) achieves this bound.

The one-dimensional harmonic oscillator has an exact cycle after some
period $\tau$.  Taking our ground-state energy to be zero, all of the
energy eigenvalues are multiples of $\varepsilon_1=h/\tau$.  Let
 \begin{equation}\eqlabel{psimax}
\ket{\psi_0} = \sum_{n=0}^{N-1} {1\over \sqrt{N}} \ket{n \varepsilon_1}
\end{equation}
 If our system passes through $N$ mutually orthogonal states in time
$\tau$, then the average time to pass between consecutive orthogonal
states is $\tau_\perp=\tau/N$.  Noting that
$\varepsilon_1\tau_\perp/\hbar=2\pi / N$, we see that the state
obtained from $\ket{\psi_0}$ after $m$ time intervals of length
$\tau_\perp$ is
 \begin{equation}
\ket{\psi_m}=\sum_{n=0}^{N-1}  {1\over \sqrt{N}} e^{-{2\pi i n m \over N}}
\ket{n\varepsilon_1}
\end{equation}
 and so
 \begin{equation}
\braket{\psi_{m'}}{\psi_m}
= \sum_{n=0}^{N-1} {1\over N} e^{{2\pi i n \over N} (m'-m)} = \delta_{m'm}
\end{equation}
  Now we can calculate the relationship between $E$ and $\tau_\perp$.
 \begin{equation}\eqlabel{e.sho}
 \bra{\psi_0}H\ket{\psi_0}
    = \varepsilon_1 \sum_{n=0}^{N-1} {n\over N}
    = \varepsilon_1 \left({N-1 \over 2}\right)
    = \left({h \over N\tau_\perp}\right) \left({N-1 \over 2}\right)
\end{equation}
  and so
 \begin{equation}
 \tau_\perp = {N-1 \over N} {h \over 2E}
\end{equation}

\subsection{Long sequences of orthogonal states}\label{longseq}

Now we turn to the question of whether ordinary macroscopic physical
systems can also run through long sequences of mutually orthogonal
states with $\tau_\perp=h/2E$.  We will show by construction that they
can.  As in the discussion above, we will not need to use arbitrarily
large eigenvalues to achieve this rate, and so our state can be
written
 \begin{equation}\eqlabel{psigen}
\ket{\psi_0} = \sum_{n=0}^{N-1} c_n \ket{E_n}
\end{equation}
 Now we simply let
 \begin{equation} c_n = \sqrt{E_{n+1} - E_n \over E_N}.
\end{equation}
 This definition of $c_n$ generalizes our example from the previous
section: for the special case of $E_n=n\ve_1$, Eq.~\eq{psigen} reduces
to Eq.~\eq{psimax}, which achieves $\tau_\perp=h/2E$ in the
macroscopic limit.  Notice also that, with this definition of $c_n$,
states with degenerate energy eigenvalues are not repeated in our
superposition (they get a coefficient of zero, since the $E_n$'s are
numbered in non-decreasing order).  This definition of $c_n$ always
gives normalized states, since
 \begin{equation}\eqlabel{norm}
\braket{\psi_0}{\psi_0} = \sum_{n=0}^{N-1} {E_{n+1}-E_n \over E_N} = 1
\end{equation}
 We can calculate the average energy in the state \ket{\psi_0}.  This is just
\begin{equation}\eqlabel{avge}
\bra{\psi_0}H\ket{\psi_0} = \sum_{n,n'}
c^*_{n'}c_n E_n\braket{E_{n'}}{E_n} 
= \sum_{n=0}^{N-1}
{E_{n+1}-E_n \over E_N} E_n
\end{equation}
 For $N\gg 1$ and $c_n \ll 1$, we can approximate this sum by an integral.
Letting $x=n/N$ and $\ve(x) = E_n/E_N$, we have
 \begin{equation}\eqlabel{longav}
\bra{\psi_0}H\ket{\psi_0} \approx E_N \int_0^1
\ve {d\ve \over dx} dx 
= {E_N \over 2} \int_0^1 {d \over dx} \ve^2 dx 
= {E_N \over 2}
\end{equation}
 In Appendix~\ref{integrals} we estimate the corrections to this
approximation, which vanish for large $N$.  Thus, with this definition
of $c_n$, by giving equal weight to equal energy intervals we get an
average energy that is half of the maximum energy, just as in the
limiting case considered in Eq.~\eq{e.sho}.

 Now the state obtained from \ket{\psi_0} after $m$ intervals of
length $\tau_\perp=h/2E=h/E_N$ is
 \begin{equation}
\ket{\psi_m} = \sum_n c_n e^{-2\pi i m {E_n \over E_N}}\ket{E_n}
\end{equation}
 and so
 \begin{eqnarray}\eqlabel{m'm}
\braket{\psi_{m'}}{\psi_m}
& = &
\sum_{n,n'} c^*_{n'}c_n e^{2\pi i{1 \over E_N}(m' E_{n'}-m E_n)}
\braket{E_{n'}}{E_n} \nonumber\\
& = & \sum_{n=0}^{N-1}
{E_{n+1}-E_n \over E_N} e^{2\pi i{E_n \over E_N}(m'-m)}
\end{eqnarray}
 As we've seen, this is equal to 1 if $m'=m$.  Let $k= m'-m \ne 0$, and
again let $x=n/N$ and $\ve(x) = E_n/E_N$.  Then
 \begin{equation}\eqlabel{m'm0}
\braket{\psi_{m'}}{\psi_m}
\approx
\int_0^1 {d\ve \over dx} e^{2\pi i k\ve} dx 
=
\int_0^1 {{d \over dx} e^{2\pi i k\ve} \over 2\pi i k} dx 
= 0
\end{equation}
 A more careful analysis (see Appendix~\ref{integrals}) verifies that
the corrections to this approximate calculation vanish for large $N$.
Thus we can run through a long sequence of nearly orthogonal states at
the rate $\nu_\perp=2E/h=E_\max/h$.

\section{Interpretation}


For an isolated macroscopic system $s$ with average energy $E^{(s)}$,
we have seen that we can construct a state that evolves at a rate
$\nu_\perp^{(s)}=2E^{(s)}/h$.  If we had many non-interacting
macroscopic subsystems, we would have an average energy for the
combined system of $E_\tot=\sum_s E^{(s)}$.  Our construction of the
previous section applies prefectly well to such a composite system,
and in particular lets us construct a state for this combination of
non-interacting subsystems that evolves at a rate of $\nu_\perp =
2E_\tot/h = \sum_s 2E^{(s)}/h = \sum_s \nu_\perp^{(s)}$.  Thus if we
subdivide our total energy between separate subsystems, the maximum
number of orthogonal states per unit time for the combined system is
just the sum of the maximum number for each subsystem taken
separately.  This is analogous to the case in a parallel computer,
where the total number of operations per second for the whole machine
is just the sum of the number of operations per second performed by
the various pieces.  Our result should be interpreted in a similar
manner: average energy tells us the maximum possible rate at which
distinct changes can occur in our system.

It is interesting to ask how this connection between energy and
maximum possible number of changes looks in a semi-classical limit.
As a simple example, let us consider a single-speed lattice
gas\cite{fhp}.  This is a classical gas model in which identical
particles are distributed on a regular lattice.  Each particle
moves from one lattice site to an adjacent site in time $\delta T$.
At the end of each $\delta T$ interval, all energy is kinetic, and all
particles have the same energy $\delta E$.  Thus if the total energy
is $E$, then $E/\delta E$ is equal to the number of particles, and so
the maximum number of spots that can change per unit of time is equal
to $2E/\delta E$: $E/\delta E$ spots can be vacated, and $E/\delta E$
new spots occupied.  Fewer spots will change if some particles move to
spots that were previously occupied, but we can never have more than
$2E/\delta E$ changes in time $\delta T$.  Thus if we impose the
constraint on this lattice system that $\delta E \delta T \ge h$, our
bound on the rate at which spots can change becomes
$\nu_{\mbox{\rm\scriptsize change}} \le 2E/\delta E \delta T \le
2E/h$.

It is also interesting to ask how this connection between energy and
orthogonal evolution looks in different inertial reference frames.
Clearly we will see some orthogonal states that are purely an artifact
of our choice of reference frame.  For example, an isolated stationary
atom in an exact energy eigenstate never transitions to an orthogonal
state, but if we view this same atom from a moving frame we will see a
sequence of distinct position states.  If we are interested in a bound
on ``useful'' dynamics (e.g., on computation), then we shouldn't count
extra states that arise just from the state of motion of the observer.
The obvious solution is to define the amount of ``useful'' dynamics to
be the number of orthogonal states that the system passes through in
its rest frame (center of momentum frame).  As long as our
non-relativistic analysis is valid in the rest frame, we can infer
that (in that frame) a system with total relativistic energy $E_r$
cannot pass through more than $2 E_r t_r /h$ different orthogonal
states in time $t_r$.  Then in any frame, we can compute this bound on
``useful'' evolution, since $E_r t_r = p_\mu x^\mu$.  In a frame in
which the system starts at the origin at time $t=0$ and moves in the
positive $x$ direction with a momentum of magnitude $p$, our bound is
$(2/h) (Et - px)$: from the time component of the bound we subtract a
space component.  Note that we subtract one orthogonal state for each
shift of a distance $h/2p$ ({\em cf.}  \cite{braunstein,yu}).

\section{Conclusion}

The average energy $E$ (above the ground state) of an isolated
physical system tells us directly the maximum number of mutually
orthogonal states that the system can pass through per unit of time.
Thus if the system is a computer, this quantity is a simple physical
measure of its maximum possible {\em processing rate}.

Equivalently we can say that $Et$ counts orthogonal states in time.
Just as accessible phase-space volume tells us the number of distinct
states that a macroscopic physical system can be put into, $Et$
``action volume'' tells us the number of distinct states that a system
with energy $E$ can pass through in time $t$.

\section*{Acknowledgments}

Margolus would like to thank Ed Fredkin, who first made him aware of
the question, ``How can we physically quantify computation.''  He and
Tom Toffoli have encouraged Margolus and inspired him in his studies
of the mechanics of information in physics.  Margolus would also like
to thank Mark Smith, Sandu Popescu, Tom Knight and Gerry Sussman for
useful comments and discussions.

Both authors would like to thank Elsag-Bailey and the
I.S.I. Foundation for the opportunity to work together on this
research at their 1996 Quantum Computing Workshop in Turin, Italy.

\appendix

\section{Minimum orthogonality time for exact cycles}\label{exact.orthog}

If we demand that the evolution should {\em exactly} cycle after $N$
steps, this condition puts a severe restriction on the energy
eigenfunctions that can contribute to our initial state.  If
$\ket{E_j}$ contributes to $\ket{\psi_0}$, and if our cycle length is
$\tau$, then we must have that $E_j \tau /\hbar = 2 \pi k_j$ for some
integer $k_j$, which means that $E_j = k_j \varepsilon_1$, where
$\varepsilon_1=h/\tau$.  Thus our initial state must have the form
 \begin{equation}
\ket{\psi_0} = \sum_{n=0}^{\infty} c_n \ket{n \varepsilon_1}
\end{equation}
 For simplicity, we have not included degenerate energy eigenfunctions
in our superposition---adding these would not affect our conclusions.
Thus the assumption of exact periodicity restricts us to systems with
energy eigenvalues that are a subset of a one dimensional harmonic
oscillator spectrum, in which all energies are multiples of $h/\tau$.

As in Section~\ref{cycles}, we see that the state obtained from
$\ket{\psi_0}$ after $m$ time intervals of length $\tau_\perp=\tau/N$
is
 \begin{equation}
\ket{\psi_m}=\sum_{n=0}^{\infty} c_n e^{-{2\pi i n m \over N}}
\ket{n\varepsilon_1}
\end{equation}
 and so
 \begin{equation}
\braket{\psi_{m'}}{\psi_m}
= \sum_{n=0}^{\infty} \av{c_n}^2 e^{{2\pi i n \over N} (m'-m)}
\end{equation}
 There are actually only $N$ distinct values of the exponential that
appears in this sum, since for any integer $l$, $e^{{2\pi i (n +lN)
\over N} (m'-m)} = e^{{2\pi i n \over N} (m'-m)}$.  We can collect
together all $c_n$'s that multiply each distinct value: let
$\av{d_n}^2=\sum_{l=0}^\infty \av{c_{n+lN}}^2$.  Then
 \begin{equation}\eqlabel{d.orthog}
\braket{\psi_{m'}}{\psi_m}
= \sum_{n=0}^{N-1} \av{d_n}^2 e^{{2\pi i n \over N} (m'-m)} 
= \delta_{m'm}
\end{equation}
 since these states are supposed to be orthogonal.  This last equality
will obviously be true if we let $\av{d_n}^2=1/N$ for all $n$'s.  In
fact, since there are $N$ different possible values of $m'-m$,
Eq.~\eq{d.orthog} constitutes $N$ linearly independent equations with
$N$ unknown coefficients, and so this solution is unique.  Thus by
picking any set of $\av{c_n}^2$'s that add up to make all the
$\av{d_n}^2$'s equal to $1/N$, we obtain a state $\ket{\psi_0}$ that
evolves at a constant rate through a sequence of $N$ mutually
orthogonal states in time $\tau$.

Now we can calculate the relationship between the average energy and
the orthogonality time.
 \begin{equation}
E=\varepsilon_1 \sum_{n=0}^\infty \av{c_n}^2 n
= \varepsilon_1 \sum_{n=0}^{N-1} \sum_{l=0}^\infty \av{c_{n+lN}}^2 (n + lN)
\end{equation}
 but $\sum_{l=0}^\infty \av{c_{n+lN}}^2 (n + lN)\geq
\sum_{l=0}^\infty \av{c_{n+lN}}^2 n = n\av{d_n}^2$, and since
$\av{d_n}^2=1/N$, 
 \begin{equation}\eqlabel{e.tau}
E \geq \varepsilon_1 \sum_{n=0}^{N-1} {n \over N} =
\varepsilon_1 \left({N-1 \over 2}\right)
\end{equation}
 Since $\varepsilon_1 = h/\tau = h/ N\tau_\perp$, this establishes
Eq.~\eq{tauofm}.  Note that we get equality in Eq.~\eq{tauofm} only
for the state given in Eq.~\eq{psimax}.

 We can also argue that if we don't use energy eigenvalues larger than
$E_\max = \varepsilon_1 (N-1)$, then there are at most $N$ eigenstates
in our superposition, and so we can pass through at most $N$ different
orthogonal states in time $\tau$.  Thus
 \begin{equation}\eqlabel{e.extra}
\tau_\perp \geq {\tau \over N} = {h \over \varepsilon_1 N} =
\left({h \over E_\max}\right) \left({N-1 \over N}\right)
\end{equation}
 which yields Eq.~\eq{maxbound} for a long exact cycle.

\section{Approximating sums}\label{integrals}

Here we estimate the corrections to the average energy and scalar
product computed in Section~\ref{longseq}.

Letting $\ve_n={E_n/E_N}$ and $\delta_n=\ve_{n+1}-\ve_n$,
Eq.~\eq{avge} becomes
 \begin{eqnarray}\eqlabel{avge2}
\bra{\psi_0}H\ket{\psi_0}& = & E_N \sum_{n=0}^{N-1}
(\ve_{n+1}-\ve_n) \ve_n \nonumber\\
& = &
{E_N\over2} \sum_{n=0}^{N-1} \left[
(\ve_{n+1}^2 - \ve_n^2) - (\ve_{n+1}-\ve_n)^2 \right]  \nonumber\\
& = &
{E_N\over2} \left[1 - \sum_{n=0}^{N-1} 
\delta_n^2 \right]  \nonumber\\
\end{eqnarray}
 Now, for any ordinary macroscopic system, from general properties of
the density of states\cite{reif} we know that for large $n$, $E_n \sim
n^c$, where $c$ is a positive constant much less that one.  From this
we can show that $\sum \delta_n^2 = O(1/N^{2c})$.

Similarly, letting $\alpha=2\pi (m'-m)=2\pi k$, Eq.~\eq{m'm} becomes
 \begin{eqnarray}\eqlabel{orthodev}
\braket{\psi_{m'}}{\psi_m}
& = &
\sum_{n=0}^{N-1}
\delta_n e^{i\alpha \ve_n} \nonumber\\
& = &
{1\over i\alpha}\sum_{n=0}^{N-1}e^{i\alpha \ve_n}\left[
(e^{i\alpha \delta_n}-1)
- \sum_{j=2}^\infty {(i\alpha\delta_n)^j \over j!}
\right] \nonumber\\
& = &
-{1\over i\alpha}\sum_{n=0}^{N-1}
 \sum_{j=2}^\infty e^{i\alpha \ve_n}{(i\alpha\delta_n)^j \over j!}
\end{eqnarray}
 Again making the assumption that $E_n \sim n^c$ for large $n$, we can
show that the magnitude of this sum is $O(k/N^{2c})$.

\end{document}